# Nonlinear chiral light generation from resonant metasurfaces

Fangxing Lai[1,2,†], Jun Yin[1†], Ivan Toftul[3†], Hamdi Barkaoui[1], Huachun Deng[1], Xinbo Sha[1], Maxim V. Gorkunov[4,5], Yuri Kivshar[3,‡], Qinghai Song[1,2,6,*]

[1]*Ministry of Industry and Information Technology Key Lab of Micro-Nano Optoelectronic Information System, Guangdong Provincial Key Laboratory of Semiconductor Optoelectronic Materials and Intelligent Photonic Systems, Harbin Institute of Technology; Shenzhen 518055, P. R. China.*

[2]*Pengcheng Laboratory; Shenzhen 518055, P. R. China.*

[3]*Nonlinear Physics Center, Research School of Physics, Australian National University; Canberra ACT 2601, Australia.*

[4]*Shubnikov Institute of Crystallography, NRC "Kurchatov Institute; Moscow 119333, Russia.*

[5]*National Research Nuclear University MEPhI (Moscow Engineering Physics Institute); Moscow 115409, Russia.*

[6]*Collaborative Innovation Center of Extreme Optics, Shanxi University; Taiyuan 030006, P. R. China.*

[†] These authors contribute equally to this research.t

Corresponding authors:
Email: ‡ yuri.kivshar@anu.edu.au, *qinghai.song@hit.edu.cn

**Abstract**

Chiral nonlinear response has been explored for decades due to its extreme sensitivity to molecular and structural dissymmetry. Conventional approaches often require bulky systems and produce only static nonlinear chirality. Here, we report on a generic mechanism for the generation and control of nonlinear chiral light in resonant optical systems. We reveal that nonlinear resonant generation of circularly polarized light from achiral dielectric metasurfaces is extremely sensitive to the polarization state of the fundamental wave, and a resonant metasurface can produce light with arbitrary degree of nonlinear chirality (DNC). Experimentally, we demonstrate that the chirality of nonlinear radiation from one metasurface can be continuously tuned from DNC = -0.86 to DNC = 0.94 by simply varying the polarization angle of the incident wave. By further exploiting the instantaneous polarization state, nonlinear chirality has been switched in a delay time step of 3.2 *fs*, which is orders of magnitude more sensitive than the current state-of-the-art polarization modulation. These results promise to enrich our understanding of nonlinear processes in chiral structures and their manipulation with resonant photonic structures.

Chiral nanophotonics is an emerging field of optics that focuses on the manipulation of light at the nanoscale with specific spin angular momentum (SAM) and the corresponding chiral light-matter interactions. (*1-5*) Over the past decade, significant efforts have been invested into this cutting-edge research area, resulting in numerous ultracompact chiral devices and their applications, such as chiral sensing (*4*), chiral transmission (*6-9*), and chiral lasing (*9*). Chiral harmonic generation from nonlinear metasurfaces is one important example of the recent developments (*3*). Due to inherent nonlinearity-induced nonreciprocity, chiroptical signals can be simply achieved from nonlinear nanostructures without breaking out-of-plane mirror-reflection symmetry, and thus they have intensively been explored in experiments. (*10-19*) Conventional approaches for achieving nonlinear response with relatively large circular dichroism mostly rely on the parametric conversion of circularly polarized light into harmonics, following certain selection rules (*12-14, 20-23*) An external quarter-wave plate is typically required for the conversion from linear to circular polarizations. By tuning the symmetry of the nanostructure, some nonlinear metasurfaces can respond only to light with specific values of SAMs (*16-19*). Severe chirality spoiling still occurs under linearly polarized excitation due to near-field disturbances between the fields with different SAMs. (*18*) Most importantly, nonlinear chirality observed in such metasurfaces is either static, or their switching time is restricted to a few picoseconds. (*24*) To the best of our knowledge, currently, there are no strategies to nonlinear generation of light with tunable chirality from achiral light using a single metasurface, especially with an ultra-sensitive response to the pump light's modulation. Here, we reveal a novel mechanism for nonlinear generation of light with arbitrary chirality and all-optical polarization control.

**Working principle of nonlinear generation of light with arbitrary chirality**

Our main concept is schematically illustrated in Fig. 1A. The achiral metasurface consists of trapezoidal Si nanopillars arranged in a square lattice embedded in a dielectric environment with a refractive index $n = 1.5$. Detailed structural parameters are depicted in the inserts of Fig. 1A. Linear transmission calculation under a *y* linearly polarized input (a solid line in Fig. 1B) shows a series of resonances in the near infrared (IR) spectral range. There are three key high Q orthogonal resonances with wavelengths of $\lambda_1 = 734.6$ nm, $\lambda_2 = 731.7$ nm, and $\lambda_3 = 731.2$ nm, and Q-factors of $Q_1 = 219$, $Q_2 = 296$, and $Q_3 = 578$, respectively. No resonances are present in the spectral range of wavelengths greater than 1 µm (Supplementary Note-1). The metasurface has both out-of-plane and in-plane mirror-reflection symmetries. Hence, both linear circular dichroism ($CD_{\text{co}} = (T_{RR} - T_{LL})/(T_{RR} + T_{LL})$) and circular conversion dichroism vanish at all wavelengths (a dashed line in Fig. 1B and Supplementary Note-1).

The metasurface exhibits completely different chiral behaviour in the same spectral range once we consider a nonlinear process. We excite the metasurface by a linearly polarized laser beam at 2196 nm and detect the degree of circular polarization at the third harmonic (TH) frequency around a 732 nm wavelength. This is different from previous reports (*3, 10-19*) that mostly utilize circularly polarized incident fields at the fundamental frequency. For a particular angle of linear polarization $\theta_1 = 83.5°$ of the incident field (left panel in Fig. 1A), the light generated from the metasurface appears to be perfectly left circularly polarized (LCP) at a 731.98 nm wavelength. The corresponding right circularly polarized (RCP) part of the radiation vanishes (marked by dash-dotted line in left panel of Fig. 1C). The quantitative characteristic of this nonlinear transformation of achiral (linearly polarized) light into chiral (circularly polarized) light can be defined as "the degree of nonlinear chirality" (DNC) following the definition $DNC =$ $(I^{3\omega}_{RCP} - I^{3\omega}_{LCP})/(I^{3\omega}_{RCP} + I^{3\omega}_{LCP})$. Here $I^{3\omega}_{RCP(LCP)}$ is the intensity of RCP (LCP) part of the TH wave generated from linearly polarized fundamental wave. In such situation, the DNC at 731.98 nm can reach -1, even though the resonances and the incident laser radiation are both achiral. Then we know that a single resonant metasurface is sufficient for directly converting linearly polarized fundamental waves into high-purity circularly polarized nonlinear signals. The external quarter waveplate for polarization conversion can thus be spared and the entire system becomes very compact.

An even more interesting phenomenon is observed when the polarization direction ($\theta$) of the fundamental wave is varied. According to the numerical simulation results shown in right panel of Fig. 1C, where the polarization direction is rotated to $\theta_2 = 96.5^\circ$ (right panel in Fig. 1A), only RCP components of the third harmonics are generated by the same metasurface, resulting in a nonlinear chirality with $DNC = 1$ at the same wavelength. Thus, the handedness of the third harmonic (TH) generation from one metasurface can be completely reversed without additional modifications on the structural parameters. To accurately capture the impact of polarization direction on nonlinear chirality, we have systematically calculated DNC as a function of the fundamental polarization direction (Fig. 1D). With the increase of polarization angel $\theta$ from $0^\circ$ to $360^\circ$ (red line), the path on the Poincare sphere (blue line) winds across both the north pole ($DNC = 1$) and the south pole ($DNC = -1$). All these observations are intrinsically different from previous reports (*3, 10-19*) and demonstrate for the very first time that the chirality of nonlinear signals from a single metasurface can be arbitrarily and post-fabrication manipulated.

To understand the origin of arbitrary nonlinear chirality generated by the same Si metasurface, we revisit the third-harmonic (TH) radiation process in nanostructures. We consider a metasurface supporting a series of eigenmodes with complex electric (magnetic) field distributions $\mathbf{e}_n$ ($\mathbf{h}_n$) and with complex angular frequency $\Omega_n - i\gamma_n$ near the THG wavelength. Assuming weak energy leakage, we normalize the eigenmodes by the factor $\sqrt{u_n/\varepsilon_0 \varepsilon V}$ where $\varepsilon$ is the relative dielectric permittivity of the embedded medium, and $u_n = \frac{1}{4}\int_V (\varepsilon(\mathbf{r})\varepsilon_0 |\mathbf{e}_n|^2 + \mu_0 |\mathbf{h}_n|^2) dV$ is the mode energy in the simulation volume $V$. The third-harmonic polarization currents are given by $\mathbf{j}^{(3\omega)} = -3i\omega\varepsilon_0 \chi^{(3)} (\mathbf{E}^{(\omega)})^2 \mathbf{E}^{(\omega)}$ and are partially coupled to the resonances. The THG field in the metasurface becomes

$$\mathbf{E}^{(3\omega)} = \sum_{n=1}^{3} a_n \mathbf{E}_n + \mathbf{E}_0^{(3\omega)}, \tag{1}$$

where $a_n$ are the dimensional amplitudes of the eigenmodes at the TH wavelength, explicitly accounted for, $\mathbf{E}_n$ are the normalized dimensionless modes, and $\mathbf{E}_0^{(3\omega)}$ is what we call the uncoupled field, i.e., the TH field generated without coupling to three explicitly accounted eigenmodes, which we refer to as implicit or nonresonant. When the TH radiation wavelength exceeds the metasurface period, no diffraction occurs. As there is no incident field at $3\omega$, the incoming radiation channels remain idle during the THG process. In our model, the metasurface thus has only four TH radiation channels — along the *x*- and *y*-directions in both forward and backward directions — with no incoming channels. The TH field radiated into these channels ($s_k$, $k$ = 1, 2, 3, 4) is obtained through a *modified* coupled-mode theory (*25*) (see Supplementary Material for derivation)

$$\begin{aligned}\frac{d}{dt}a_n &= -i(\Omega_n - i\gamma_n)a_n - \Gamma \sum_{k=1}^{4} d_{kn}^* s_k^{(0)} - \frac{1}{4\varepsilon_0 \varepsilon V}\int_V \mathbf{E}_n^* \cdot \mathbf{j}^{3\omega} dV \\ s_k &= \sum_{n=1}^{3} d_{kn} a_n + s_k^{(0)}\end{aligned} \tag{2}$$

The latter overlap integral in the first line describes the driving of the resonant mode by the third-harmonic current density. In Eq. (2), $\Gamma = v/2L$ is a constant with the dimension of inverse time, where $v$ and $L$ are the speed of light in the embedding medium and the height of the modeling box, respectively. We stress that final result does not depend on the choice of the integration volume as long as it contains the resonator. The coefficient $d_{kn}$ characterizes the radiation efficiency of the $n$-th mode into the $k$-th outgoing channel. The term $s_k^{(0)}$ represents the contribution to the THG from the nonresonant uncoupled component of the third-harmonic field $\mathbf{E}_0^{(3\omega)}$.

Following Eqs. 1, 2 and eigenmode expansion method (*27*), the TH radiation in the simulation can be decomposed to several resonant modes and four uncoupled fields. Their contributions to DNC are then precisely identified. One example is depicted in Fig. 2A. The radiation behavior at 731.98 nm in Fig. 1B is expanded to *x*- and *y*- components of the forward radiation of the uncoupled field $\mathbf{E}_0^{(3\omega)}$ and the main component of the forward radiation of three eigenmodes contribution, whose field patterns are shown in Fig. 2B. The amplitudes and phases of such components are strongly correlated and vary significantly with the polarization state of the fundamental waves. The total radiation at $\theta = 83.5°$, the vector sum of all these components, amounts to an electric field with equal and orthogonal real and imaginary parts providing a circularly polarized output. Over time, the magnitude of this electric field remains constant, and its direction rotates clockwise (Fig. 2C), consistent with the LCP TH emissions. The amplitudes of all components at $\theta = 96.5°$ are identical to the ones at $\theta = 83.5°$ (two vertical dashed-dotted lines in Fig. 2A). Nonetheless, the jump of some phases completely reverses the rotation direction of the vector sum (Fig. 2D), resulting in a pure RCP THG. At other polarization angles, the amplitudes and phases vary simultaneously and thus the corresponding polarization states wind a dramatic path on the surface of Poincare sphere across the south pole and north pole (dots in Fig. 1D).

**Experimental demonstration of THG with arbitrary chirality**

Based on the above simulation and analysis, we have fabricated Si metasurfaces on K9 glass substrate using a combined process of electron-beam lithography and inductively coupled plasma etching (Supplementary Note-3). (*29*) Figure 3A shows the top-view scanning electron microscope (SEM) image of one Si nanostructure. The detailed structural parameters can be seen from the high-resolution SEM images in the inserts and are summarized in Supplementary Note-3. The Si metasurfaces are obtained after coating the nanostructures with a 500 nm polymethyl methacrylate (PMMA) film. The linear transmission spectrum is measured (Supplementary Note-3) and plotted as a solid line in Fig. 3B. Several resonances can be clearly seen, including the designed dip at around 734.6 nm due to three resonant modes. The transmission spectra with different handedness have also been characterized (Supplementary Note-3). The corresponding linear CD is calculated and plotted as dashed line in Fig. 3B. It remains at CD ≈ 0 in the entire spectral range, consistent with the mirror-reflection symmetries of metasurface. The slight deviation from zero quantifies experimental measurement errors.

Then the metasurface is optically excited by a femtosecond laser at 2220 nm (pulse width 100 fs, repetition rate 1 kHz) and the nonlinear process is explored (Fig. 3C). (*30*) Similar to the numerical simulation, here the laser is linearly polarized at an angle $\theta = 45º$ to the *x*-direction and no external quarter waveplate is used (Supplementary Note-3). The light in the forward direction is analyzed using a spectrometer after passing a short-pass filter. An emission peak rises at around 740 nm and a red spot appears on the sample. The corresponding THG conversion efficiency is $1.7 \times 10^{-7}$. Figure 3D shows the integrated intensity of emission peak as a function of pumping fluence far below the damage threshold (161 mJ/cm$^2$). A slope of 2.982 has been obtained in the log-log plot. All these experimental results demonstrate clearly that THG process happens in the Si metasurface. The polarization state of the TH signals has also been analyzed by placing a linear polarizer and a quarter waveplate behind the short pass filter. Although the metasurface is achiral and the fundamental wave is linearly polarized, the LCP output is obviously larger than its counterpart with opposite SAM, giving nonlinear chirality with DNC = -0.75 (Supplementary Note-4). Therefore, we confirm that a single resonant metasurface has the capability to directly convert linearly polarized waves at the fundamental frequency into nonlinear chiral light at the corresponding harmonic.

One of the remarkable features of our mechanism is the generation of TH with arbitrary nonlinear chirality from the same metasurface. Experimentally, we explore this potential by gradually rotating the polarization angle of fundamental wave from $\theta = 0°$ to $\theta = 180°$. Figure 3E summarizes the dependence of nonlinear chirality on the linear polarization angle at 735.41nm with the dispersive silicon (See *n_Si.txt*). The nonlinear chirality of THG vanishes at $\theta = 0°$, and it decreases gradually with the rotation of the polarization angle. The degree of chirality reaches its minimal value of DNC = -0.86 at $\theta = 60°$, where the TH signal is dominated by the LCP component (top panel in Fig. 3F). DNC rises from the minimum by further increasing the polarization angle. For the case of $\theta = 120°$, the TH signals are almost entirely RCP (bottom panel in Fig. 3F), and the nonlinear chirality reaches a maximal value of DNC = 0.94. It can be observed that nonlinear chirality has been precisely controlled and reversed between the values DNC = -0.86 to DNC = 0.94 by a simple rotation of the incident polarization (see details in Supplementary Note-4). All the above observations are consistent with our numerical results (dashed line in Fig. 3E, details in Supplementary Note-1) and confirm our mechanism well. Note that the refractive indices of PMMA and K9 glass are slightly different in experiments. Our numerical simulation shows that the contribution of such a tiny out-of-plane asymmetry to DNC of THG can be neglected (see Supplementary Note-1).

**All-optical control of THG chirality**

In principle, the arbitrary chirality is attributed to the complexity of nonlinear polarization states. As demonstrated above, the chirality of nonlinear signals depends strongly on the polarization states of the fundamental waves, e.g., the polarization direction and ellipticity. In this sense, the instantaneous polarization state within an ultrashort laser pulse can be employed to all-optically tune the nonlinear chirality. (*23, 31*) Typically, an optical pulse with orthogonal components ($E_x(t)$ and $E_y(t)$) has a time-varying polarization state if $E_x(t)$ and $E_y(t)$ are not strictly proportional (Supplementary Note-5). Such an instantaneous polarization state holds true even for a single-cycle pulse. As a result, polarization transitions on the sub-*fs* scale become possible for the fundamental waves. The same can be expected for the corresponding nonlinear chirality, only restricted by the fundamental pulse (Supplementary Note-6).

In the experiment, we achieve the instantaneous polarization state with a combination of delay line and liquid crystal retarder (Fig. 4A and Supplementary Note-5). (*23, 32-35*) One beam is elliptically polarized with $2\Psi = 2.44$ and $2\chi = 0.64$, where the other beam is linearly polarized along $\theta = 180°$. The power ratio between two beams is set as $|A_1|^2 = 1.2|A_2|^2$. DNC for the third harmonics generated by two individual beams is 0.20 and -0.01, respectively. When the delay time $\Delta\tau$ is greater than 100 fs, the nonlinear chirality of the entire THG is thus averaged out, and the corresponding nonlinear chirality vanishes at the designed wavelength (top panel in Fig. 4B). The case for a delay time below 100 *fs* is more involved. As shown schematically in the insets of Fig. 4A, the size (A), direction ($2\Psi$), and ellipticity ($2\chi$) of polarization ellipse vary dramatically within the overlap of the two beams, producing instantaneous polarization states. The instantaneous DNC transition is hard to measure experimentally. Interestingly, the average THG of all polarization states at each delay time also changes significantly and can be characterized. Our theoretical simulation reveals that the DNC of THG repeats quasi-regularly with a period of 7.36 *fs* (green line in Fig. 4C). This is exactly what we have observed in the experiment. As the dots depicted in Fig. 4C, the averaged nonlinear chirality varies and switches between the values 0.5 and -0.5 with a quasi-period of around 7.3 *fs* (Supplementary Note-6). Figure 4D shows the enlarged transition within one period, where DNC can be switched from 0.47 to -0.46 ultra-sensitively within an ultrashort 3.2 *fs* delay time difference between two pump pulses. The corresponding THG spectra are plotted in middle panel and bottom panel of Fig. 4B. It can be seen that the THG helicity is reversed at a delay time step being orders of magnitude shorter than that reported previously (*24, 32-40*) and the conventional resonant shift (Supplementary Note-6).

In experiments, system stability affects the results significantly. In our experiment, we excluded this possibility by conducting three measurements on the same samples (Supplementary Note-6). All three measurements show quite similar trends with tiny deviations. The statistical analysis shows that the averaged periodicities and transitions from DNC = 0.5 to DNC = -0.5 are almost identical and consistent with the numerical simulations, clearly demonstrating the stability of our system well. For simplicity, we focused above only on the resonance near the TH wavelength. It can also be extended to other spectral ranges as well, e.g., for telecommunication wavelengths (Supplementary Note-7).

In summary, we have proposed and verified experimentally a universal quantitative physical mechanism of nonlinear generation of chiral light using a metasurface and a pumping wave which are both achiral (The whole system is a chiral system). The process allows for all-optical control: we have demonstrated that the degree of nonlinear chirality of THG from the same metasurface can cover a wide range between the values -0.86 and 0.94, and, for the first time to our knowledge, the nonlinear chiral response can be switched and tuned ultra-sensitively within an ultrashort delay time step around 3.2 *fs*. We believe that these findings will enrich our understanding of nonlinear chiral metaphotonics, driving a paradigm shift in information processing and chiral sensing (Supplementary Note-8).

## Methods

**Numerical simulations**

The linear properties are obtained with finite element method. Periodic boundary conditions are applied in *x*- and *y*- directions to simulate the periodic nanostructures. Perfect matching layers (PML) are used in vertical directions to absorb the outgoing waves and mimic the infinite size of outer space. The transmission spectrum under linear polarization has been simulated by calculating the energy ratio of the transmitted wave to incident at each wavelength. A two-step approach is implemented in the frequency domain simulation of THG. The metasurface is excited with a plane wave at the fundamental frequency $\omega$, and the induced nonlinear polarization is then calculated from the fundamental field pattern. Subsequently, the third-order polarization generated from the fundamental frequency is utilized as a source to excite the metasurface at the THG frequency and accurately simulate the THG radiation. The time-domain THG simulation with pulses input is performed in FDTD with similar boundary conditions as those in the linear simulations. The Gaussian pulse is incident downward onto the structure and silicon is set as the Kerr nonlinear material in FDTD. We perform a Fourier transformation on the mean electric field in the XY plane approximately 675 nm below the bottom surface of the trapezoidal pillar to obtain the frequency spectra of the electric field components along the *x*- and *y*-directions. These spectra are subsequently converted into LCP and RCP field spectra.

**Fabrication of Si metasurfaces**

The Si metasurface is fabricated as follows. Amorphous Si film with a thickness of 234 nm (Syskey, 0.5 Å/s) is deposited onto a K9 glass substrate and then coated with a 21 nm thick Cr film (Syskey, 0.3 Å/s) by E-beam evaporation. The sample is then covered by a 100 nm thick E-beam resist (PMMA A2) by spin-coating at a speed of 4000 r/min. After baking on a hot plate at 180 °C for an hour, the sample is patterned with an E-beam aligner (Raith E-line) and inversed pattern is generated in E-beam resist by developing in MIBK:IPA for a ratio of 1:3 for 30 seconds. The sample is dried and further coated by a 15 nm thick $SiO_2$ film via E-beam evaporation (Syskey, 0.4 Å/s). The designed pattern is realized in $SiO_2$ layer after a lift-off process. Taking the $SiO_2$ patterns as a mask, two steps of inductively coupled plasma processes are employed to etch the Cr and Si layers. The residuals of Cr and $SiO_2$ on it are removed by a striping process and the sample is coated by a PMMA layer via spin-coating under negative pressure.

**Optical characterization**

Supercontinuum laser is used to characterize the linear transmission spectra and the corresponding CD of the Si metasurface. Linear polarization is generated with a linear polarizer and can be further converted to circular polarization by using a quarter-wave plate. After passing through a quarter-wave plate and a linear polarizer, the corresponding left-handed and right-handed circular components of the transmitted light can be recorded and by a spectrometer. Femtosecond laser with a center wavelength of about 2220 nm and pulse length of about 100fs is then introduced into the optical setup for THG experiments. The linear polarization state of femtosecond laser is adjusted by a half-wave plate and a linear polarizer. It is then focused onto the metasurface through a 10× objective lens (NA = 0.25) and collected by a 20× objective lens (NA = 0.4). All-optical control of nonlinear chirality is realized with a pump-probe system, the pump fluence of one beam ($2\chi = 0.64$) is 5.0 mJ/cm$^2$ and the pump fluence of another beam ($2\chi = 0$) is 4.1 mJ/cm$^2$. These two pulses are also focused onto the metasurface through a 10x objective lens (NA = 0.25) and the output THG is collected by a 20x objective lens (NA = 0.4). By adjusting the position of translation stage and phase of liquid crystal retarder, the relative delay time between two pulses can be accurately controlled and the instantaneous polarization state of incident wave on the metasurface is tuned accordingly. The polarization state and spectral information of THG are analyzed with the same process of the linear ones.

## Acknowledgments

This work is supported by National Key R&D Program of China (Grant Nos. 2021YFA1400802, 2022YFA1404700), National Natural Science Foundation of China (Grant Nos. 6233000076, 12334016, 11934012, 12025402, 62125501, 12261131500, and 92250302), Shenzhen Fundamental Research Project (Grant Nos. GXWD20220817145518001), Fundamental Research Funds for the Central Universities (Grant Nos. 2022FRRK030004, 2023FRFK03049), Australian Research Council (Grant No. DP210101292), International Technology Center Indo-Pacific (ITC IPAC) via Army Research Office (contract FA520923C0023), New Cornerstone Science Foundation through the XPLORER PRIZE., and the joint Russian-Chinese project supported by the Russian Science Foundation (Grant nos. 23-42-00091).

## Author contributions

Q.S. and Y.K. conceived the idea and supervised the project. F.L., Q.S. designed the experiments. F.L., J.Y., I.T., H.B. conducted simulations. J.Y., I.T., M.G., Y.K., Q.S. built the theoretical model. F.L., H.D., X.S. performed the experiments. F.L., Q.S. analyzed the data. Q.S. and Y.K. drafted the paper with extensive input from all authors.

**Competing interests**

Authors declare that they have no competing interests.

**Data and materials availability**

All data are available in the main text or the supplementary materials. (1)

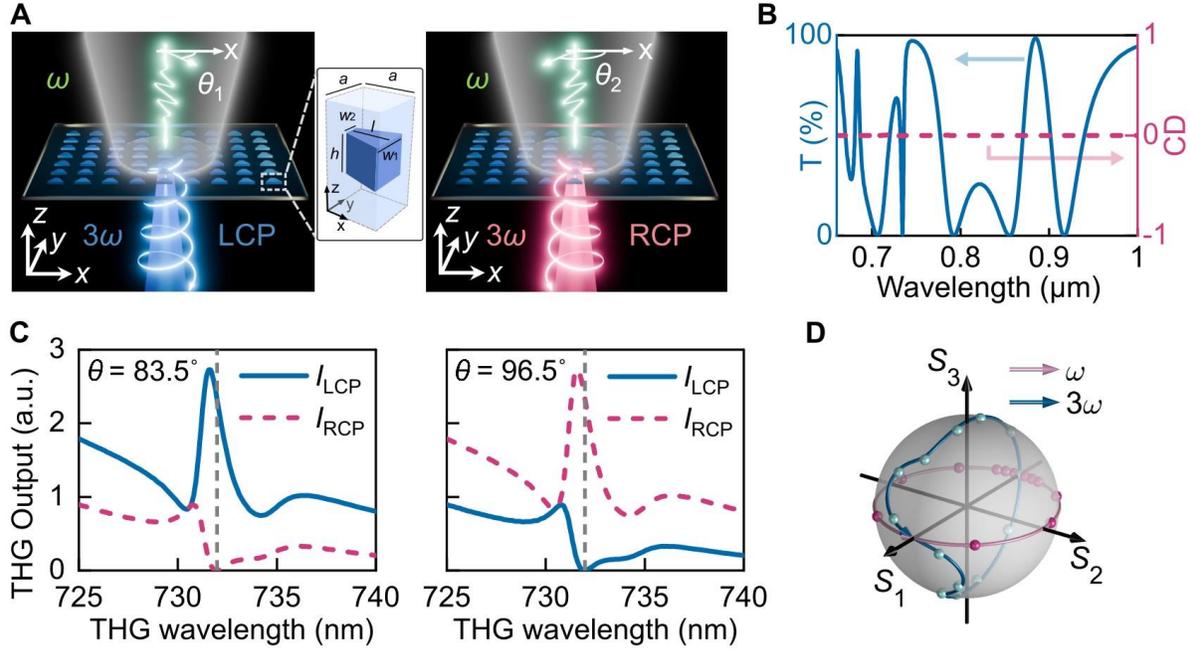

**Figure 1: Nonlinear radiation with arbitrary chirality generated by a resonant metasurface.**
**A**. Schematics for the generation of LCP (left panel) and RCP (right panel) THG with the same Si metasurface. The reversal of THG handedness is achieved with a simple change in polarization direction ($\theta$). Inserts show the three-dimensional schematic diagram of one unit cell with structural parameters. **B**. Linear transmission under the *y*-linearly polarized input (solid line) and the corresponding CD (dashed line) of the Si metasurface. **C**. shows the numerically calculated LCP (solid line) and RCP (dashed line) THG spectra with different polarization angles of $\theta = 83.5°$ and $96.5°$, respectively. The resonant wavelength is marked by a vertical dash-dotted line. **D**. The winding path of numerically calculated THG chirality (solid line) on the surface of Poincare sphere as a function of polarization angle of fundamental waves. The dots represent the fitting results with the theoretical model. Here the size parameters are $a = 434$ nm, $w_1 = 379$ nm, $w_2 = 89$ nm, $l = 160$ nm, and $h = 234$ nm. The refractive index of Si is n = $3.9246 - 9.978 \times 10^{-4}i$.

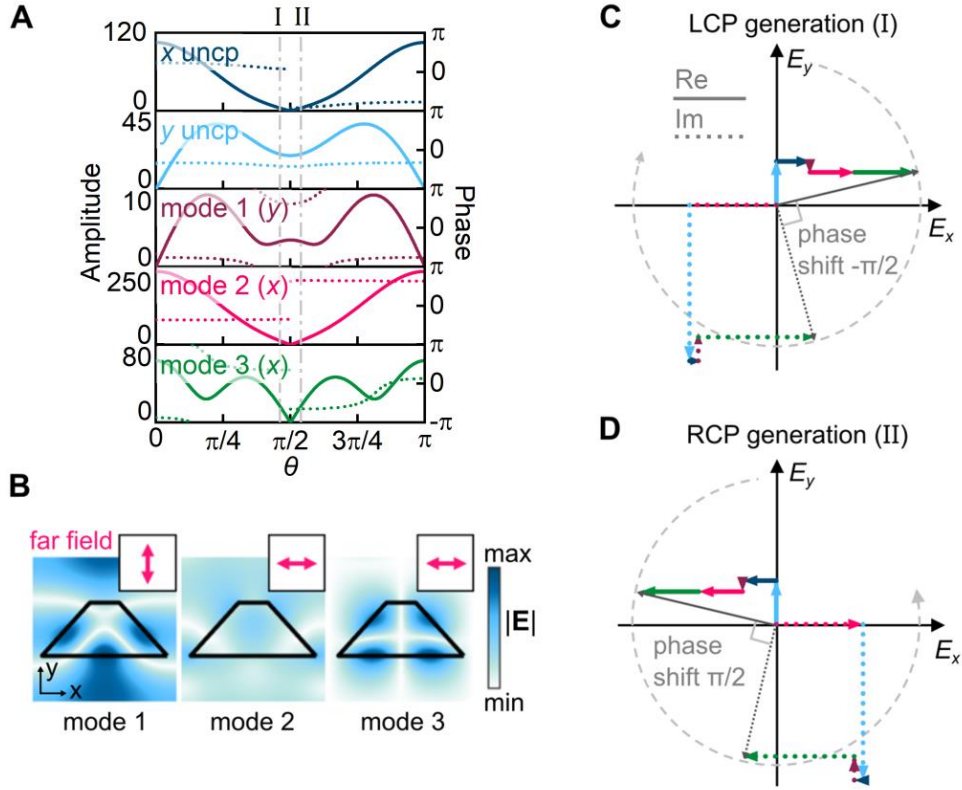

**Figure 2: Working principle for arbitrary chirality of TH radiation. A**. The amplitudes (solid lines) and phases (dashed lines) of the *x*- and *y*-components of the superposition of forward TH radiation induced by the *x* and *y* components of the uncoupled field $\mathbf{E}_0^{(3\omega)}$ "*x* uncp" and "*y* uncp" (top two panels), and the amplitudes (solid lines) and phases (dashed lines) of the main component (*y* component for mode 1, and *x* component for mode 2 and 3) of the forward radiation of three expanded eigenmodes (bottom 3 panels) at the TH wavelength 731.98 nm, which are excited differently at different polarization angle $\theta$. **B**. The electric field amplitude (|*E*|) of three eigenmodes at 734.6 nm, 731.7 nm, and 731.2 nm, respectively. The fields are taken from the *x-y* plane in the middle of trapezoid nanopillar, and their corresponding forward radiation polarization are shown in the top-right boxes. **C** and **D** are the real (gray solid arrow) and imaginary (gray dotted arrow) parts of vector sum of all components at $\theta = 83.5°$ and $\theta = 96.5°$, respectively, marked by the corresponding vertical dashed-dotted lines I and II in A. All expanded components are plotted in the same colors as in A. Their individual real and imaginary parts are also denoted as solid arrows and dotted arrows, respectively. A slight change in the incident polarization state leads to a complete reversal of the THG helicity in C and D.

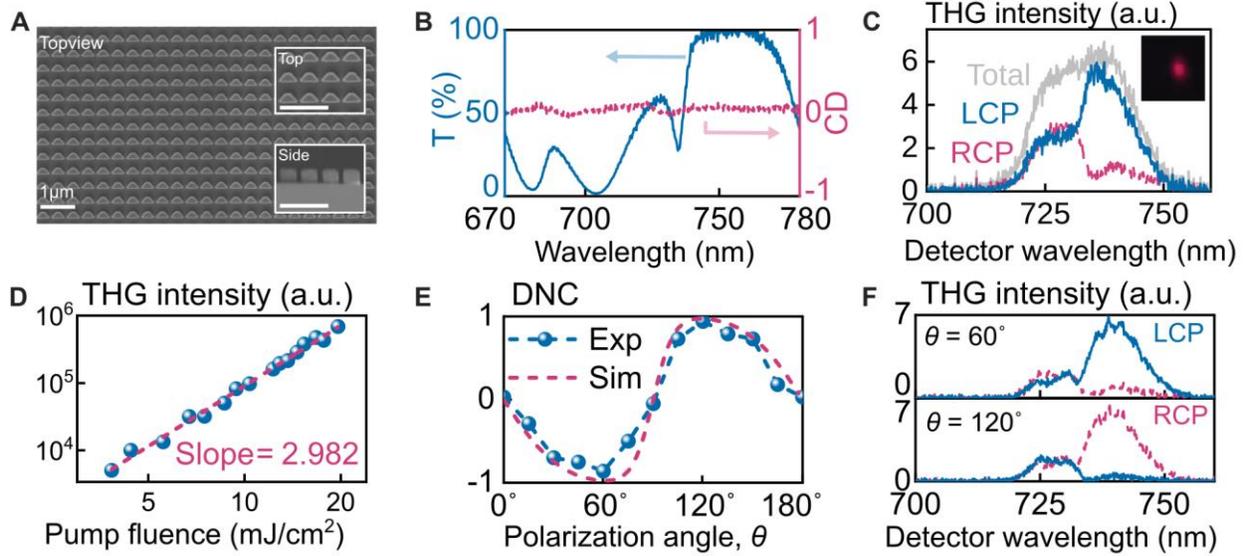

**Figure 3: Experimental demonstration of arbitrary chiral THG from a metasurface.** **A**. Top-view SEM image of Si metasurface. The detailed structural parameters can be seen from high-resolution top-view (top inset) and side-view (bottom inset) SEM images. The scale bars are 1 *μm*. **B**. Experimentally recorded transmission spectra (solid line) and the corresponding linear CD (dashed line). **C**. The experimentally recorded spectra (gray line) from the metasurface and the corresponding near-field microscope image (inset). Blue solid line and red dashed line show the LCP and RCP components of the emission. Here the polarization angle $\theta$ is 45º. **D**. The integrated output as a function of input power. **E**. The dependence of DNC of THG on the polarization angle $\theta$ of fundamental waves at 735.41nm with the dispersive silicon (See *n_Si.txt*). The numerical results are plotted as dashed lines for a direct comparison. **F**. The LCP and RCP intensity of THG at the polarization angle of $\theta = 60º$ (top panel) and $\theta = 120º$ (bottom panel), respectively. The excitation energy density in C, E and F is 13.2 mJ/cm². The metasurface is excited by a femtosecond laser at 2220 nm.

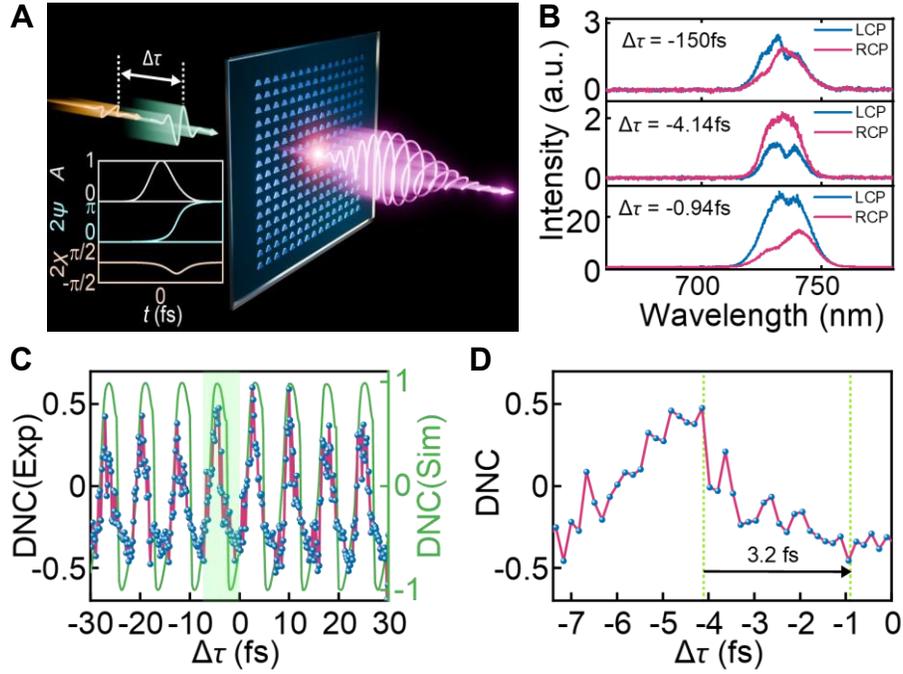

**Figure 4: All-optical control of nonlinear chirality in Si metasurface. A**. Schematic of the all-optical nonlinear chirality control. Two pulses with different polarization states and a tunable time delay are employed to control the time-variable polarization state. Inserts show the size (top panel), orientation (2ψ, middle panel), and ellipticity (2χ, bottom panel) of the superposition of two pulses with Δτ= 50 *fs*. **B**. The LCP and RCP THG spectra at Δτ= -150 *fs* (top panel), Δτ= -4.14 *fs* (middle panel), and -0.94 *fs* (bottom panel), respectively. **C**. Comparison between experimental data (dots and red line) and simulated data (green solid line) of DNC as a function of delay time Δτ. **D**. Enlarged figure for an ultra-sensitive transition of nonlinear chirality under an ultrashort delay time modulation of the pump light. DNC is dramatically switched with an ultrashort delay time step of 3.2 *fs*.